# THE IMPACT OF FILL PATTERNS ON THE FAST ION INSTABILITY IN THE ILC DAMPING RING

Guoxing Xia, School of Physics and Astronomy, the University of Manchester, Manchester, U.K.,
The Cockcroft Institute, Daresbury, U.K.

*Abstract*

The ions produced via collisional ionization of the residual gas molecules in vacuum pipe with the circulating electron beam have deleterious effect on the beam properties and may become a limiting factor for the machine's performance. For the electron damping ring of the International Linear Collider (ILC), the ion instability is noticeable due to the ultra-low beam emittance with many bunches operation. In this paper, the different beam fill patterns are investigated and their effects on the fast ion instability are discussed. The simulations show that the mini train fill patterns can reduce the growth of the fast ion instability significantly.

## INTRODUCTION

The ionization of the residual gas in the vacuum pipe by the circulating electron beam will create positive ions which will under certain circumstances become trapped in the potential well of the stored beam [1,2]. The accumulation of these ions depends upon several factors, e.g. the number of bunches, bunch spacing, the beam current (or bunch populations), transverse beam sizes (or beam emittances and the machine optics) and the mass of the trapped ions etc. In general, the accumulation of these ions is detrimental to the machine's performance. It may lead to beam tune shift and spread, emittance blowup, the coupling of horizontal and vertical motion, the electron bunch ion cloud oscillations etc.

The damping rings for the International Linear Collider (ILC) play an important role to achieve the ultra-low emittance beams. The lattice design of the ILC damping rings has evolved significantly since the first ILC baseline design report was issued in 2005. Until 2009, a new design, so-called Strawman Baseline or SB2009 has been worked out based on the idea to optimize the cost and performance of the ring according to the ILC Reference Design Report (RDR issued in 2007). The criterion to choose a damping ring design is based on the ring's performance, e.g. the large dynamic aperture and acceptance (especially for positron beams), electron cloud and fast ion instability thresholds, and also on the fast kicker development etc. In 2011, a new ILC damping ring lattice DTC02 has been designed [3]. It features a racetrack structure with a circumference of 3.2 km. This design includes two arc sections consisting of the Theoretical Minimum Emittance (TME) cells and two long straight sections consisting of FODO cells. In the straight section, the injection/extraction system, RF, wigglers, chicane for circumference tunning and phase trombone are located. The basic beam parameters of the ring are listed in Table 1.

Table 1: Beam parameters for the DTC02 damping ring.

| Energy (GeV) | 5.0 |
|---|---|
| Circumference (m) | 3238.68 |
| Harmonic number | 7022 |
| Bunch number | 1312-2625 |
| Bunch spacing (buckets) | 4-2 |
| Number of particles/bunch | $1\text{-}2\times10^{10}$ |
| Damping time $\tau_x$ (ms) | 24 |
| Normalized emittance $\varepsilon_x$ (μm) | 3.8 |
| Momentum compaction $\alpha$ | $1.07\times10^{-4}$ |
| Synchrotron tune | 0.059 |
| Energy spread | $1.21\times10^{-3}$ |
| Bunch length (mm) | 6 |
| RF frequency (MHz) | 650 |
| RF voltage (MV) | 5-16 |

For most electron storage rings, the produced ions are accumulated over many turns and trapped in the beam potential. This beam-ion cloud instability can be partially cured by either applying electrostatic clearing electrodes, beam shaking or intentionally leaving some RF buckets empty in the fill pattern. However, in order to get a high luminosity of $2\times10^{34}$ cm$^{-2}$s$^{-1}$, the ILC damping rings need to provide a large number of bunches (bunch number ranges from 1312 to 2625 bunches and therefore the bunch spacing is narrow) with ultra-low beam emittance (vertical emittance ~2 pm) electron and positron bunches for the downstream main linacs and the beam delivery system. The bunch intensity is $1\sim2\times10^{10}$. In this case, the ions produced by a single passage of electron bunches may become significant. This so-called fast ion instability (FII) will cause beam size growth and beam emittance blowup and therefore it is detrimental to the damping ring's performance [4, 5]. In this paper, we investigate the fast ion instability based on the latest DTC02 damping ring design. Different fill patterns and their effects on the growth of fast ion instability are investigated.

## SIMULATION STRATEGY

A weak-strong simulation code is used in our study [6]. In the simulation, the beam is represented by a rigid Gaussian marco-particle. The beam sizes of bunches are therefore fixed (we also assume that the beam is already damped to low emittance) and only their dipole motions are investigated and computed every turn. The ions are generated at positions of all optical elements in the ring lattice. New marco-particles for the generated ions are produced at the transverse position $(x, x', y, y')$ of the beam where the ionization occurs. The number of ions is increased with respect to the number of bunches in the bunch train. The ion line density per bunch is given by $\lambda_{ion} = N_0 \sigma_{ion} n_{gas}$, here $N_0$ is the bunch population, $\sigma_{ion}$ is the ionization cross section of gas and $n_{gas}$ is the gas density. In the simulation, we assume that the first electron bunch in the bunch train produces the ions and it does not interact with the ions. The following bunches interact with the ions produced by their preceding bunches. After one turn interaction, the ions are cleared away from the beam centre. The new batch of ions will be produced by the beam in the second turn. To save the simulation time, we use only one arc section of DTC02 lattice in which the beta functions and dispersion function vary at different locations along the ring. In these interaction points (each element in the lattice corresponds to an interaction point), we artificially enhance the number of ions by taking into account the real vacuum pressure of the ring. The adjacent beam-ion interaction points are connected through the linear transfer matrix.

We consider that the main ion species in the vacuum chamber are Carbon Monoxide ($CO^+$) and Hydrogen ($H_2^+$) [7]. The cross section of collision ionization for CO is about 6 times higher than that for $H_2$ in the beam energy of 5 GeV. In addition, the heavy mass of CO makes it easier to be trapped in the beam potential compared to the Hydrogen. We therefore consider $CO^+$ ions the dominant instability source in our simulation.

## SIMULATION RESULTS

In order to get a high luminosity, the ILC features a flat beam operation, that is to say, the vertical beam emittance is much smaller than the horizontal one, therefore the FII is much severer in the vertical plane. In our simulations, the time evolution of the beam dipole amplitude is simulated and recorded turn by turn. The data are recorded for 1000 turns which is about a half damping time. The vertical oscillation amplitude of the bunch centroid is half of the Courant-Synder invariant and given by

$$J_y = [\gamma y^2 + 2\alpha y y' + \beta y'^2]/2$$

where $\alpha$, $\beta$ and $\gamma$ are the Twiss parameters of the ring determined by the ring lattice (MAD output file). The value of $\sqrt{J_y}$ is compared with the vertical beam size which is represented by the value of $\sqrt{\varepsilon_y}$ (here $\varepsilon_y$ is the beam vertical emittance).

Fig. 1 shows the beam maximum vertical oscillation amplitude due to FII with respect to the number of turns for a single long bunch operation in which 1312 bunches are filled in the ring. In this figure, $N_0$ denotes the number of particles per bunch, $n_t$ is the total bunches in the beam and $S_b$ is the bunch spacing. We assume the CO partial pressure of 1.0 nTorr. It can be seen that the beam oscillation amplitude is above the beam size (vertical beam size is around $1.4\times10^{-6}$ m, as depicted by a dot line in the figure). The growth time of FII is also estimated and shown in the Fig.1. We can see that in this case the FII growth time is about 6 turns, which is too fast for a feedback system to cure it. A typical damping time of the fast bunch-by-bunch feedback is about 0.2 ms from the experience of KEKB [8], which corresponds to ~20 turns for the ILC damping rings. Fig.2 gives the beam maximum oscillation amplitude with respect to the number of turns with the same beam parameters as used in Fig.1 at CO pressure of 0.1 nTorr. It can be seen that the FII growth time is around 39 turns, which is longer than that at CO pressure of 1 nTorr due to fewer ions produced at lower gas pressure.

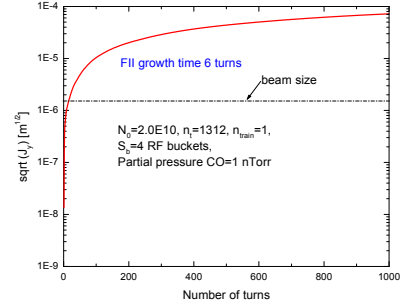

Figure 1: Beam oscillation amplitude *vs.* number of turns for 1312 bunches in a single train at 1.0 nTorr CO.

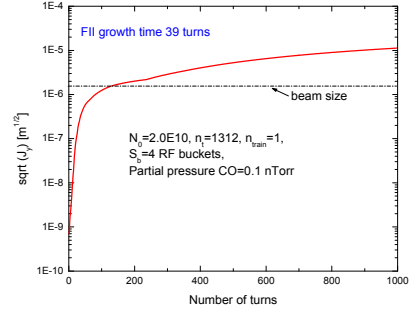

Figure 2: Beam oscillation amplitude *vs.* number of turns for 1312 bunches in a single train at 0.1 nTorr CO.

We know that the empty RF buckets (gap) can over focus the ions in the beam, making them drift away from the beam centre and therefore alleviating the ion instability [9,10]. We also simulate this scenario by dividing a long bunch train into many mini-trains (or sub-trains), and with some empty RF buckets in between. Fig.3 shows the beam oscillation amplitude versus number of turns for 41 bunch trains ($n_{train}$), with each consisting of 32 bunches ($L_{train}$) at 1.0 nTorr CO. 43 RF buckets ($L_{trainGap}$) are left for ions clearing. The result shows that in this case, the FII growth time is about 26 turns which is longer compared to a long bunch train case in Fig.1. If we introduce the feedback system with a damping time of 20 turns for this fill pattern, as shown in Fig. 4, the FII growth is greatly reduced and the beam maximum oscillation amplitude is below the beam size.

Figure 5 gives the beam oscillation amplitude with respect to the number of turns for 2625 bunches in a single bunch train at CO gas pressure of 0.1 nTorr. The bunch intensity is $1.0\times10^{10}$. The growth time of FII in this case is about 27 turns. Similarly, if we divide this long bunch train into 35 mini-trains (50 RF buckets as gap), with each consisting of 75 bunches and also apply the fast feedback system, as shown in Fig.6, the growth of FII is reduced and the beam centroid oscillation amplitude is well below the beam size at CO pressure of 1.0 nTorr. Fig.7 shows another typical fill pattern in which 2625 bunches are divided into 75 trains, with each consisting of 35 bunches (shorter bunch train compared to that shown in Fig.6). 23 RF buckets are left for ion clearing in this case. The result shows that at CO pressure of 1.0 nTorr, the growth of FII is well below the beam size. We also find that the growth of FII is even smaller compared to that shown in Fig.6.

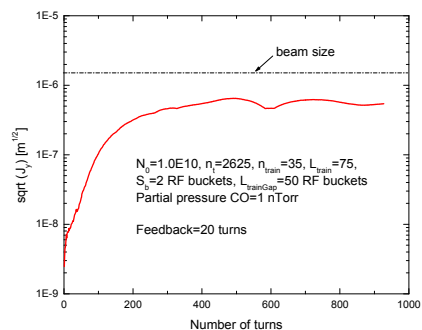

Figure 6: Beam oscillation amplitude *vs*. number of turns for 35 bunch trains, with each consisting of 75 bunches at 1.0 nTorr CO, with a feedback damping time of 20 turns.

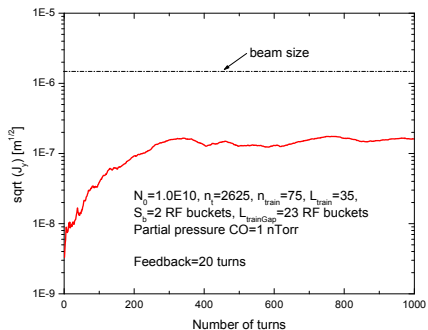

Figure 7: Beam oscillation amplitude *vs*. number of turns for 75 bunch trains, with each consisting of 35 bunches at 1.0 nTorr CO, with a feedback damping time of 20 turns.

## CONCLUSION

An extensive study of the fast ion instability based on the DTC02 lattice has been done and reported in this paper. The conclusion is that, the FII grows very fast for a single bunch train operation at CO pressure of 1 nTorr. The mini-train fill pattern, in which a long bunch train is divided by many relatively short trains, and with the gap in between, proves to be a relatively simple and effective way to eliminate the growth of fast ion instability. For the ILC electron damping ring, we can control the fast ion instability through combining the mini-train fill patterns and the fast feedback system with a damping time of ~ 20 turns.

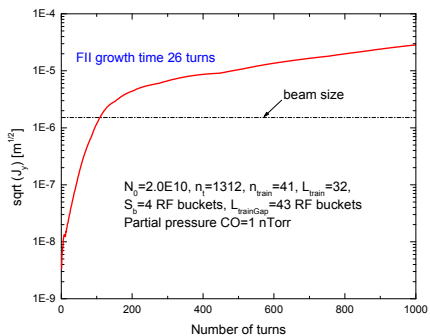

Figure 3: Beam oscillation amplitude *vs*. number of turns for 41 bunch trains, with each consisting of 32 bunches at 1.0 nTorr CO.

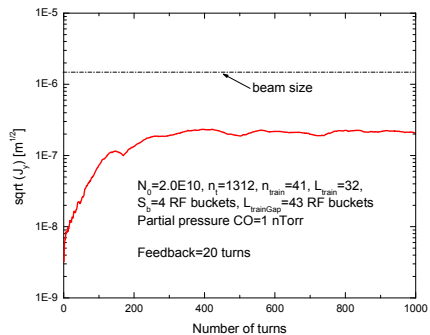

Figure 4: Beam oscillation amplitude *vs*. number of turns for 41 bunch trains, with each consisting of 32 bunches at 1.0 nTorr CO, with a feedback damping time of 20 turns.

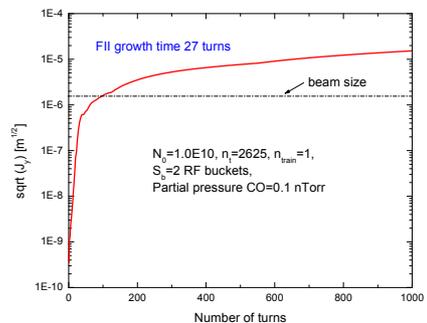

Figure 5: Beam oscillation amplitude *vs*. number of turns for 2625 bunches in a single train at 0.1 nTorr CO.